# Adaptive Meta-Aggregation Federated Learning for Intrusion Detection in Heterogeneous Internet of Things

Saadat Izadi[a], Mahmood Ahmadi[*a]

[a] Department of Computer Engineering and Information Technology, Razi University, Kermanshah, Iran
* Corresponding author.

E-mail addresses: s.izadi@razi.ac.ir (Saadat Izadi), m.ahmadi@razi.ac.ir (Mahmood Ahmadi).

**Abstract**

The rapid proliferation of the Internet of Things (IoT) has brought remarkable advancements to industries by enabling interconnected systems and intelligent automation. However, this exponential growth has also introduced significant security vulnerabilities, making IoT networks increasingly targets for sophisticated cyberattacks. The heterogeneity of IoT devices poses critical challenges for traditional intrusion detection systems. To address these challenges, this paper proposes an innovative method called Adaptive Meta-Aggregation Federated Learning (AMAFed), designed to enhance intrusion detection in heterogeneous IoT networks. By employing a dynamic weighting mechanism using meta-learning, AMAFed assigns adaptive importance to local models based on their data quality and contributions, enabling personalized yet collaborative learning across devices. The proposed method was evaluated on three benchmark IoT datasets: ToN-IoT, N-BaIoT, and BoT-IoT, representing diverse real-world scenarios. Experimental results demonstrate that AMAFed achieves detection accuracy up to 99.8% on ToN-IoT, with F1-scores exceeding 98% across all datasets. On the N-BaIoT dataset, it reaches 99.88% accuracy, and on BoT-IoT, it achieves 98.12% accuracy, consistently outperforming state-of-the-art approaches.

## 1. Introduction

The rapid advancement of Internet of Things (IoT) technology has significantly expanded its applications across diverse domains, but this progress has also intensified security concerns. The growing number of attacks targeting physical devices, combined with the emergence of unknown and sophisticated cyber threats, has exposed IoT environments to unprecedented vulnerabilities. Many of these devices are resource-constrained and interconnected, creating a large attack surface that adversaries can exploit. As a result, the development of intelligent and adaptive security mechanisms capable of addressing evolving threats has become imperative. For decades, Intrusion Detection Systems (IDSs) have played a vital role in identifying malicious activities within networks [1–3].

Recently, Federated Learning (FL) [4–6] has been introduced as a decentralized paradigm that enables collaborative model training without transferring raw data, thereby mitigating privacy concerns and reducing communication of sensitive information. However, despite its privacy-preserving nature, FL is not immune to security risks, such as model poisoning where compromised clients submit manipulated updates to degrade detection performance, and inference attacks, which can reveal sensitive patterns from shared model parameters. These risks highlight the necessity of employing robust aggregation strategies that can resist adversarial or low-quality updates in FL-based IDS frameworks [7–9]. Most existing FL-based IDS approaches, however, make simplifying assumptions that limit their practicality. Many rely on homogeneous or synthetic datasets, which do not reflect the diversity and complexity of real-world IoT environments. This mismatch often leads to degraded accuracy and poor generalization when deployed in heterogeneous IoT (HetIoT) networks. Moreover, common aggregation methods such as FedAvg treat all client updates equally, disregarding variations in data quality, local performance, and trustworthiness [10]. Prior works [11–13] have explored the challenges of applying FL to IDS, but often without providing detailed evaluations or addressing heterogeneity in depth.

The rapid growth of HetIoT expected to surpass 17 billion devices by 2025 [14–18] has introduced a wide range of malicious activities, including Distributed Denial of Service (DDoS) attacks, one of the most disruptive threats [15]. These environments are characterized by extreme heterogeneity in device capabilities, traffic patterns, and attack distributions. Such diversity complicates the task of distinguishing between normal and malicious behaviors, often resulting in increased false alarm rates and reduced detection accuracy [19–21]. IDSs in these contexts must effectively learn from varied, imbalanced, and dynamically changing data sources. However, heterogeneity can hinder FL model convergence and even cause system collapse when client updates conflict or fail to represent the global distribution. Some studies have examined FL aggregation under heterogeneous conditions [22–24], and meta-learning has emerged as a promising approach to assign dynamic weights to client updates based on their quality, diversity, and historical performance [25, 26]. By tailoring aggregation to the characteristics of client data, meta-learning can improve global model generalization, accelerate convergence, and mitigate overfitting particularly in the presence of data heterogeneity.

To address these challenges, this work introduces Adaptive Meta-Aggregation Federated Learning (AMAFed), a federated IDS framework designed for heterogeneous IoT environments. AMAFed employs a meta-learning-based aggregation strategy that dynamically adjusts client contributions according to data characteristics and local performance metrics. Additionally, a hybrid loss function combining cross-entropy and Dice coefficient is adopted to improve sensitivity to rare attack types, while an anomaly-aware regularization term emphasizes contributions from clients with critical or underrepresented attack patterns. This design enables the system to adapt to evolving data distributions, resist malicious updates, and maintain robust detection performance. The proposed method is evaluated on three benchmark datasets ToN-IoT [27], N-BaIoT [28], and BoT-IoT [29] representing diverse real-world IoT scenarios, achieving notable improvements in accuracy, precision, recall, and F1-score. Thus, the objective of this work is to design a scalable, secure, and adaptive FL-based IDS that addresses the dual challenges of data heterogeneity and adversarial client behavior, thereby enhancing intrusion detection effectiveness in real-world HetIoT environments. The main contributions of this study are as follows:

- **Adaptive Meta-Aggregator–** A novel aggregation mechanism that dynamically assigns client weights based on data distribution and local performance, effectively handling data heterogeneity and improving convergence.
- **Hybrid Loss Function–** An integrated loss combining Cross-Entropy and Dice coefficient to address class imbalance and enhance the detection of rare attack patterns without compromising accuracy on frequent classes.
- **Anomaly-Aware Regularization –** A global regularization strategy that prioritizes updates from clients containing underrepresented but critical attack patterns, strengthening the model's resilience to heterogeneous data and improving generalization in real-world IoT environments.

The paper is structured as follows: A survey of relevant works in the area of FL-based intrusion detection in the Internet of Things is provided in Section 2. Section 3 describes the suggested method, while Section 4 discusses the evaluation's findings and contrasts it with earlier methods. The conclusion and next steps are finally discussed in Section 5.

## 2. Related Works

In recent years, the development of IDS with FL capabilities in the IoT scenario has attracted attention due to its features and strengths [30]. This section provides an overview of recent studies conducted in this field. FL was introduced as a decentralized learning approach to ensure privacy preservation. In the paper referenced as [21], a thorough and comprehensive review is presented, focusing on the application of FL in IDS. The review covers challenges, different aspects of intrusion detection, and offers valuable insights into novel ideas. Additionally, Farahi et al. [31] conducted a comprehensive study utilizing FL techniques for cybersecurity in IoT applications. Furthermore, the vulnerabilities in security and privacy systems based on FL have been investigated. In this study, three different deep learning methods, specifically Recurrent Neural Network (RNN), Convolutional Neural Network (CNN), and Deep Neural Network (DNN), were employed to assess and analyze their effectiveness. The performance of both centralized and FL was evaluated for each model using three authentic datasets of IoT traffic, namely Bot-IoT, MQTTset and ToN-IoT dataset. In [32], a review of the application of FL in IDS has been presented, highlighting the existing limitations in recent works and proposing future directions. Li et al. [33] have proposed a combined approach called A FL Empowered Architecture, based on FLand fog/edge computing, to combat DDoS attacks. The proposed method aims to train an optimized global model by leveraging distributed datasets, thereby addressing data and communication limitations. However, the authors did not provide specific details regarding performance metrics such as the number of clients involved or the number of training epochs conducted. In a related study, Gampas et al. [22] evaluated an IDS approach enabled by FL. Their evaluation considered a multi-class classifier and took into account various data distributions to detect various attacks in an IoT scenario. They have also considered the impact of different aggregation functions in their work.

Hey et al. [34] introduced a federated-based approach for intrusion detection that utilized synthetic data and unsuitable configurations. In [11], they primarily addressed the challenges and future prospects of IDS with FL capabilities. However, they did not include evaluation results for their proposed methods nor establish a set of metrics for comparing the impacts within the IoT field. Liu et al. [35] briefly discussed FL in their research and put forward a FL-based approach for intrusion detection. They classified and categorized two types of attacks, namely poisoning and inference attacks. In [36] proposed a privacy-preserving scheme based on FL called VerifyNet. To ensure privacy, they utilized smooth encoding, pseudo-random technology, and a double-masking protocol using homomorphic encryption. Zhao et al. [37] have designed an IDS that utilizes FL to detect vulnerable IoT devices. Their approach involves sharing an initial global short-term memory model among all user servers. Afterwards, each user server generates its own distinct model and transmits the model configurations to the central server. According to the provided information, the central server collects and combines the model configurations from the user servers to create a new comprehensive model. This comprehensive model is then transmitted back to the central server. Mothukuri et al. [12] have introduced an anomaly detection approach based on FL to actively detect intrusions in IoT networks, with a focus on addressing privacy concerns associated with centralized machine learning methods. Experimental findings indicate that their proposed approach surpasses classical/centralized machine learning (non-FL) approaches in terms of preserving user data privacy and achieving optimal performance.

Chen et al. [38] have presented a novel approach called FDAGMM, which is an automatic deep federated Gaussian mixture model specifically designed for anomaly a centralized framework, revealing a notable enhancement in the accuracy of the federated approach for attack detection. Another study proposed a FL-based framework for intrusion detection in IoT [39], which ensures data privacy by performing local training and preserving the inference models for detection. Liu et al. [40] provides a comprehensive overview of the integration of Federated Learning (FL) and Meta-Learning for addressing cybersecurity challenges. The authors focus on how combining these two techniques can enhance the performance of Intrusion Detection Systems (IDS) in distributed and privacy-sensitive environments, particularly in the context of IoT networks. The challenge of data heterogeneity in IoT environments, particularly when employing Federated Learning (FL), has been a significant focus in recent research.

An innovative method dubbed FDAGMM, an automatic deep federated Gaussian mixture model created especially for anomaly in a centralized framework, was introduced by Chen et al. [38]. They found that the federated technique for attack detection significantly improved accuracy. Another study put out a FL-based architecture for IoT intrusion detection [39], which protects data privacy by executing local training and maintaining the detection inference models. Liu et al. [40] provides a comprehensive overview of the integration of Federated Learning (FL) and Meta-Learning for addressing cybersecurity challenges. The authors focus on how combining these two techniques can enhance the performance of Intrusion Detection Systems (IDS) in distributed and privacy-sensitive environments, particularly in the context of IoT networks. The challenge of data heterogeneity in IoT environments, particularly when employing Federated Learning (FL), has been a significant focus in recent research. To address this issue, this paper proposes a Meta-Aggregator model based on federated learning that effectively manages data heterogeneity by dynamically adjusting the aggregation function based on the characteristics of each client's local data. This dynamic aggregation process improves the performance of intrusion detection systems (IDS) in IoT networks by reducing false positives and negatives, and enhancing the overall robustness of the model. Table 1 summarizes the methods, advantages, and limitations of the studies mentioned in the text.

## 3. Proposed Method

In this section, the problem is first defined, and then the proposed method will be presented.

### 3.1. Problem Statement

The rapid growth of Internet of Things (IoT) networks has led to significant security concerns, particularly regarding the detection of intrusions across diversity and heterogeneity. This heterogeneity complicates the development of reliable Intrusion Detection Systems (IDS) as it leads to challenges in accurately identifying malicious activities. The significance of addressing this problem lies in the increasing frequency and sophistication of cyber-attacks targeting IoT networks, which can have devastating consequences if left undetected. Traditional intrusion detection methods, while effective in homogeneous environments, struggle to handle the complexity and variability of IoT systems, making it essential to explore new approaches for improving IDS accuracy in such settings. Previous research has introduced Federated Learning (FL) as a promising solution to mitigate privacy concerns and overcome the limitations of centralized machine learning models in IoT networks. FL enables decentralized learning, where local models are trained on individual devices, and only model updates are shared with a central server. However, most of the existing FL-based IDS methods primarily focus on homogeneous data environments, where the data characteristics across devices are relatively uniform. Few studies have addressed the challenges of applying FL to heterogeneous IoT networks, where differences in device capabilities and attack types pose unique obstacles for model aggregation and generalization.

Table 1: Summary of related works for FL-based IDS

| | **Method Proposed** | **Advantages** | **Limitations** |
|---|---|---|---|
| Mothukuri et al. [12] | FL-based anomaly detection for IoT networks, focusing on privacy concerns. | Outperforms centralized ML methods in privacy preservation and performance. | Lacks exploration of heterogeneous IoT environments. |
| Agrawal et al. [21] | Comprehensive review on FL in IDS, discussing challenges and offering insights into novel ideas. | Broad overview of FL's role in IDS, highlighting challenges and opportunities. | Does not propose or evaluate a specific method; no experimental validation. |
| Campos et al. [22] | Evaluated FL-enabled IDS with multi-class classification and varying data distributions. | Explored the impact of different aggregation functions; detected various IoT attacks. | Limited scope in aggregation analysis; no real-world deployment scenario. |
| Koroniotis et al. [28] | FL Empowered Architecture combining FL with fog/edge computing to tackle DDoS attacks. | Addresses data and communication limitations; proposes a distributed approach for model training. | Lacks detailed performance metrics, client information, and training specifics. |
| Belenguer et al. [32] | Review of FL applications in IDS, focusing on limitations and proposing future directions. | Identifies limitations and gaps in FL-based IDS research. | No experimental results or concrete solutions provided. |
| Hei et al. [34] | FL-based approach for IDS focusing on poisoning and inference attacks. | Addressed specific attacks in FL; categorized attacks effectively. | Limited discussion on overall IDS performance and data heterogeneity. |
| Lyu et al. [35] | Explored FL techniques in IoT cybersecurity using RNN, CNN, and DNN, and assessed centralized vs. FL-based methods. | Multi-model comparison; evaluated FL's effectiveness on diverse datasets (Bot-IoT, MQTTset, ToN-IoT). | Limited to comparison; lacks innovation in FL aggregation techniques. |
| Chen et al. [38] | FDAGMM: FL-based Gaussian Mixture Model for attack detection in IoT. | Achieves higher accuracy in attack detection; efficient anomaly identification. | Limited scalability analysis; requires centralized initialization. |
| Rahman et al. [39] | Federated-based IDS using synthetic data. | Highlights FL challenges and prospects in IDS. | Lacks evaluation results and comparison metrics; unsuitable configurations. |
| Liu et al. [40] | Integration of FL and Meta-Learning for cybersecurity challenges in IoT networks. | Combines two advanced techniques to enhance IDS performance; targets distributed environments. | Lacks detailed evaluation results; limited to theoretical insights. |
| Proposed Method | Adaptive Meta-Aggregation Federated Learning (AMAFed) | Dynamically adjusts model aggregation based on data characteristics, improving detection in heterogeneous IoT environments | Increased computational complexity |

The primary challenges in this context include data heterogeneity, the complexity of varying attack types, and the insufficient data for less common attacks. These factors contribute to a higher risk of false alarms and reduced detection accuracy in traditional FL-based IDS models. Additionally, the standard aggregation techniques used in FL, such as FedAvg, often fail to account for the quality and relevance of data from diverse sources, leading to ineffective model convergence and compromised performance. As IoT networks continue to expand, these issues are expected to worsen, further highlighting the need for innovative solutions. The goal of this research is to propose an enhanced FL-based IDS framework capable of effectively addressing the challenges posed by data heterogeneity in IoT environments. This study aims to improve the aggregation process by leveraging dynamic weighting mechanisms and meta-learning to adapt the contribution of each local model based on data quality and relevance. By doing so, the research intends to develop a more accurate and robust intrusion detection system that can better cope with the diverse and evolving nature of IoT networks.

*3.2. Proposed solution*

The proposed intrusion detection system for heterogeneous IoT environments utilizes a multi-layered architecture consisting of Edge, Fog, and Cloud layers, as illustrated in Figure 1. Initially, the central server at the Cloud layer initializes a global model with random parameters $\theta_g$ and distributes it to all k clients (IoT devices) across the Fog layers. Each client k has a unique local dataset $D_k$ that may differ significantly in terms of data distribution and quality.

At the Fog layer, the intrusion detection model training proceeds as follows:

- **Local Training:** Each client k receives the global model $\theta_g$ and trains it on its local dataset $D_k$, resulting in updated local model parameters $\theta_k$.
- **Adaptive Weight Calculation:** Upon completing local training, client kk computes an adaptive weight $w_k$ that quantifies the quality and relevance of its local data and model. This weight is dynamically calculated using meta-features extracted from the local dataset and model performance metrics. Specifically, data diversity is measured by the distribution of attack types present in $D_k$, indicating how representative and varied the client's data is. In addition, model performance is evaluated through metrics such as the false positive rate (FPR) on local validation data, which assess the accuracy and reliability of $\theta_k$. These meta-features are combined into a scoring function to produce $w_k$, ensuring that clients with more diverse and higher-quality data, or better local model performance, receive higher weights.
- **Transmission to Server:** Each client sends its updated model parameters $\theta_k$ along with the adaptive weight $w_k$ to the central server.
- **Meta-Aggregation at Server:** The central server performs a weighted aggregation of all received client models to update the global model $M_g$, where each $\theta_k$ is scaled by its corresponding weight $w_k$. This approach prioritizes updates from clients whose data and models are more informative and reliable, effectively addressing the heterogeneity of data across the network.
- **Weight Update Dynamics:** The adaptive weights $w_k$ are recalculated at each training round (epoch) to reflect the most recent data characteristics and model performance. This dynamic recalculation enables the system to adapt to changing local data distributions and client behavior throughout the training process. If certain clients experience data drift or degraded model quality, their weights naturally decrease, reducing their influence on the global model.
- **Iteration Until Convergence:** The updated global model $M_g$ is redistributed to all clients for the next round of local training. This cycle of local training, adaptive weight calculation, and weighted aggregation repeats until the global model converges, ensuring it generalizes well across the heterogeneous IoT environment. To summarize the key parameters and their roles, Table 2 provides detailed definitions.

Table 2: Model Parameters and their descriptions.

| Parameter | Symbol | Description | Role in the model |
|---|---|---|---|
| Global model parameters | $\theta_g$ | Initial model parameters initialized by the central server. | Serve as the starting point for training at each client. |
| Local dataset | $D_k$ | The unique dataset available to each client k, exhibiting potentially heterogeneous data. | Provides training data for local model updates. |
| Local model parameters | $\theta_k$ | Updated parameters obtained after training the global model on $D_k$ | Reflect local data characteristics in the model. |
| Adaptive weight | $w_k$ | Weight computed by each client based on data quality, diversity, and local performance metrics. | Determines the influence of the client's model on the global aggregation. |
| Global model | $M_g$ | The global model obtained after aggregating local updates using adaptive weights $w_k$ | Represents the optimized intrusion detection model for all IoT devices. |
| Meta-objective | $\Phi$ | A function that calculates client weights based on data characteristics and performance metrics. | Responsible for computing adaptive weights ($w_k$) for each client in each training round. |
| Data distribution coefficient | $\gamma$ | Controls the influence of data distribution metrics (attack types) on the client weighting function. | Ensures that clients with diverse or underrepresented data distributions are prioritized. |
| Model performance coefficient | $\delta$ | Controls the influence of model performance metrics (e.g., accuracy, false positives/negatives) on the client weighting function. | Ensures that clients with higher-quality local models contribute more to the global model. |
| Learning rate | $\eta$ | Controls the step size for each weight adjustment during the gradient-based optimization process. | Ensures smooth and stable updates to the client weights ($w_k$), preventing over/under-adjustment. |
| Regularization weight | $\alpha$ | Weight of the regularization term $R(\theta)$ in the global loss function. Controls the influence of overfitting prevention and model generalization. | Balances the effect of $R(\theta)$ to avoid overfitting and ensure the model generalizes effectively. |
| Anomaly-aware weight | $\beta$ | Weight of the anomaly-aware regularization term $A(\theta)$ in the global loss function. Balances sensitivity to rare or unique attack patterns. | Adjusts the model's focus on rare attack patterns to improve detection in heterogeneous datasets. |

The proposed method, adaptive meta-aggregation federated learning, addresses data heterogeneity in IoT intrusion detection through its adaptive meta-aggregation approach, the general architecture of which is presented below. The proposed Meta-Aggregator addresses data heterogeneity in intrusion detection across diverse IoT devices by dynamically adjusting the influence of each client based on specific data characteristics. This adaptive approach is achieved through three main mechanisms. First, clients that detect rare or high-risk attack types are assigned greater weights, ensuring these critical patterns are well-represented in the global model, thereby enhancing the model's sensitivity to diverse and uncommon attack types. Second, clients with a balanced distribution of attack and normal data receive prioritized weighting, reducing the model's bias toward majority data patterns and ensuring a more accurate detection of both common and infrequent attack scenarios. Finally, at each federated learning round, the meta-objective $\Phi$ recalculates client weights based on updated meta-features and performance metrics, allowing the model to adaptively respond to changing data distributions among clients. This dynamic weighting system strengthens the model's robustness and generalization, enabling reliable intrusion detection in complex, heterogeneous IoT environments. However, the proposed method is systematically organized into several components, which are detailed in the following subsection (Figure 2).

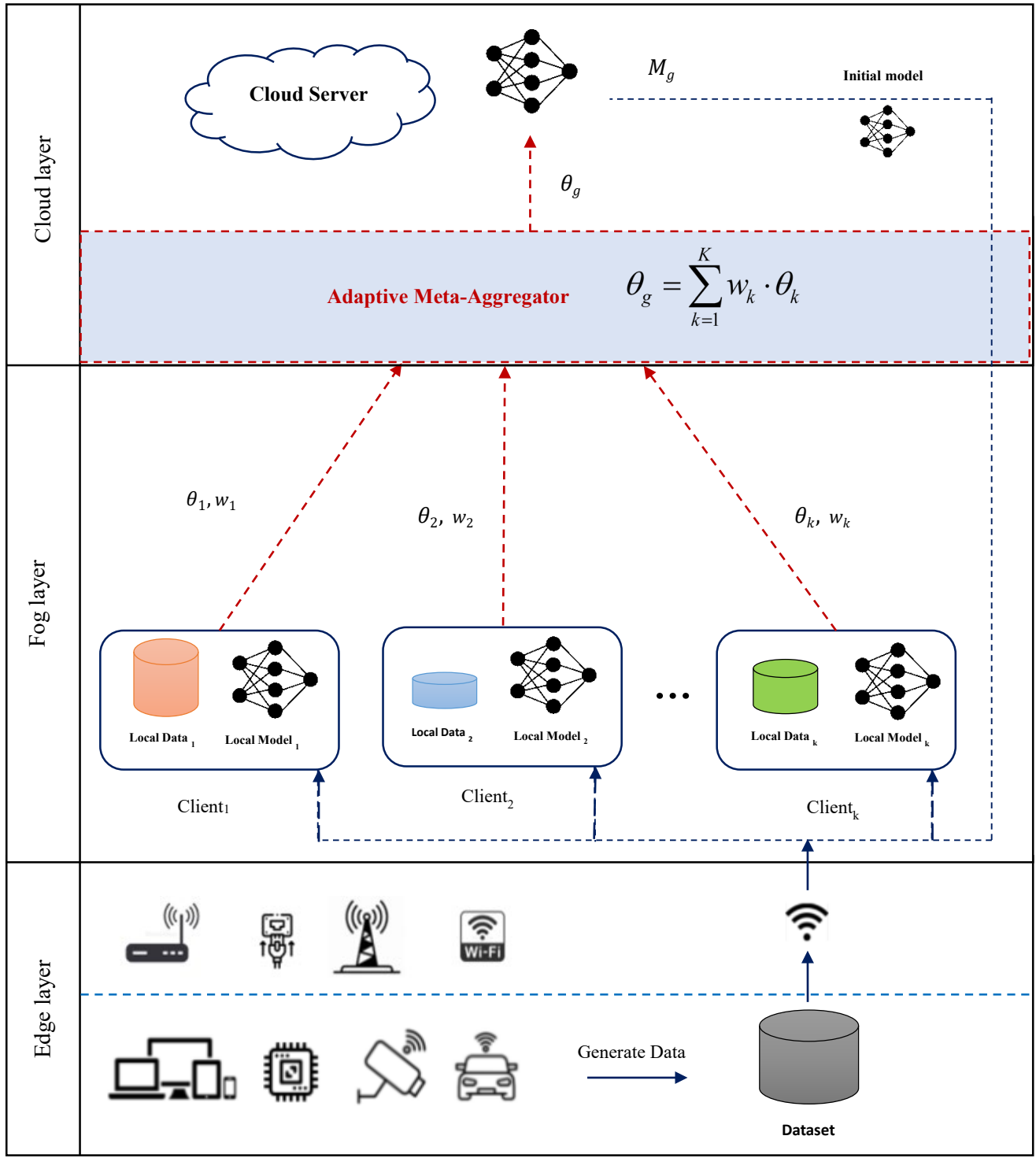

Figure 1: Architecture of IDS in HetIoT based on proposed method

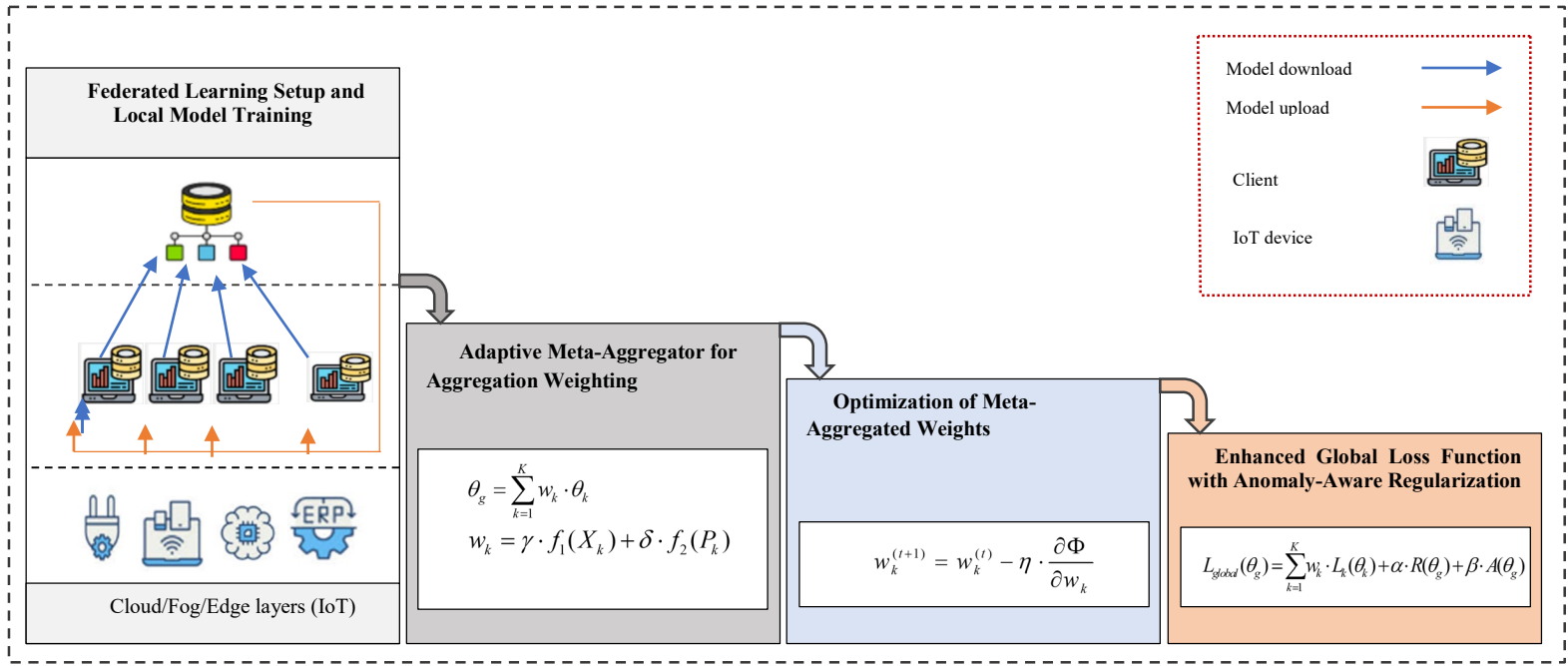

Figure 2: Process of proposed method

Each subsection systematically elaborates on a specific aspect of the proposed method, providing a comprehensive understanding of its structure and innovations.

- **Federated learning setup and local model training**: this subsection describes the initial configuration of the federated learning framework, including the setup and training process of local models on edge devices.
- **Adaptive meta-aggregator for aggregation weighting**: the meta-aggregation mechanism, designed to dynamically adjust aggregation weights based on device heterogeneity, is elaborated upon in this part.
- **Optimization of meta-aggregated weights**: the optimization process for refining the meta-aggregated weights, ensuring an optimal balance between local contributions and global objectives, is discussed in detail.
- **Enhanced global loss function with anomaly-aware regularization**: this section introduces a specialized loss function incorporating anomaly-aware regularization to enhance the accuracy of intrusion detection.

### 3.2.1. Federated Learning Setup and Local Model Training

In the FL setup, K clients (IoT devices) are considered where each client $k$ has: a local dataset $D_k$ and local model $M_k$ with parameters $\theta_k$. The objective is to collaboratively build a global model $M_g$ with parameters $\theta_g$, which can generalize effectively across all clients. Each client $k$ minimizes its local loss function $L_k(\theta_k)$, which evaluates the model performance on its local dataset $D_k$ (Eq. 1).

$$L_k(\theta_k) = \lambda \cdot \text{CE}(\hat{y}_k, y_k) + (1 - \lambda) \cdot \text{Dice}(\hat{y}_k, y_k) \quad (1)$$

Where, **CE**: cross-Entropy, a loss function commonly used for classification, **Dice**: dice coefficient, which measures the similarity between the predicted and actual data distributions, **λ:** a hyperparameter that balances the contribution of CE and Dice (λ typically ranges between 0 and 1), and $y_k$ and $\hat{y}_k$ represent actual output value and predicted output value, respectively. The Cross-Entropy Loss (CE) measures the difference between predicted and actual probability distributions in classification tasks, while the Dice Coefficient evaluates segmentation accuracy by measuring the overlap between predicted and actual binary masks, where N is the total number of samples, C is the number of classes, and k represents each sample index (Eq. 2).

$$CE(y, \hat{y}) = -\frac{1}{N}\sum_{k=1}^{N}\sum_{c=1}^{C} y_{k,c} \cdot \log(\hat{y}_{k,c}), \quad Dice(y, \hat{y}) = \frac{2\sum_{k=1}^{N} y_k \cdot \hat{y}_k}{\sum_{k=1}^{N} y_k + \sum_{k=1}^{N} \hat{y}_k} \qquad (2)$$

To address the difference in scale between the CE and Dice loss functions, which normalize each loss before combining them. This normalization ensures that both losses contribute proportionally, preventing one from dominating the optimization process. Specifically, it uses min-max normalization to scale each loss to the range [0,1]. The combination of cross-entropy and the dice coefficient is crucial for enhancing detection accuracy in imbalanced datasets. By adjusting λ, it can prioritize one loss function over the other, depending on the characteristics of the dataset. This dual approach increases the model's sensitivity to rare attack instances while preserving overall classification performance. By leveraging both metrics, it can effectively minimize the risk of false negatives.

*3.2.2. Adaptive Meta-Aggregator for Aggregation Weighting*

To effectively address data heterogeneity, this approach introduces an adaptive meta-aggregator that dynamically assigns a weight $\omega_k$ to each client's model, leveraging data-specific meta-features and performance metrics. These adaptive weights are recalculated at the end of each training epoch to reflect the most recent data distribution and model performance of each client. Each client's weight $\omega_k$ is calculated based on key meta-features, including: data distribution metrics (such as the proportion of malicious to normal packets and the types of attacks detected) and model performance metrics (like local model accuracy, and rates of false positives and false negatives). The global model $M_g$ is then derived through an optimization process that strategically balances each client's contribution, creating a more resilient model that better handles the complexity and variability inherent in IoT environments (Eq. 3) [41].

$$\theta_g = \sum_{k=1}^{K} w_k \cdot \theta_k \qquad (3)$$

where $w_k$ represents the optimal weight assigned to each client's model parameters. This weighting function adapts dynamically, taking into account the heterogeneity of each client's data. To compute the weight $w_k$, a weighted combination of these metrics is used. Specifically, $w_k$ is calculated as follows:

$$w_k = \gamma \cdot f_1(X_k) + \delta \cdot f_2(P_k) \qquad (4)$$

Where, $f_1(X_k)$ is a function based on data distribution metrics such as the ratio of malicious to normal packets and the types of attacks detected, $f_2(P_k)$ is a function based on model performance metrics such as accuracy and false positive/negative rates, and $\gamma$ and $\delta$ are coefficients that control the influence of data distribution and model performance on the final weight. This weighting mechanism allows the model to prioritize clients with high-quality data and strong local performance, thus enhancing the generalization ability of the global model. The exact values of $\gamma$ and $\delta$ can be determined through experimentation or optimization techniques, ensuring that the weighting function effectively balances the contribution of each client. This weight recalculation occurs after each local training phase, so if the client's data characteristics or model quality change over time (for example, due to concept drift or new attack types emerging), the system automatically adjusts the influence of that client in the global aggregation. In other words, the adaptive weights are not fixed once computed but evolve dynamically with each training round, ensuring the global model continuously prioritizes clients contributing the most relevant and high-quality information. Finally, the global model parameters $M_g$ are updated using these dynamic weights via weighted aggregation (Eq.5).

$$\theta_g = \frac{\sum_{k=1}^{K} w_k \cdot \theta_k}{\sum_{k=1}^{K} w_k} \qquad (5)$$

This iterative and dynamic weighting mechanism enhances the robustness and generalization ability of the federated intrusion detection system across heterogeneous IoT data.

*3.2.3. Optimization of Meta-Aggregated Weights*

To compute adaptive weights $w_k$ for each client, a meta-optimization objective $\Phi$ is formulated, designed to prioritize clients based on data quality and diversity. The optimization goal is to maximize contributions from clients with data that enriches the model's ability to generalize effectively across the network. This optimization problem is structured as follows (Eq. 6) [41].

$$\min_{w_k} \Phi(w_k, X_k, P_k) \quad \text{subject to} \quad \sum_{k=1}^{K} w_k = 1, \quad w_k \geq 0 \qquad (6)$$

where: $X_k$ : meta-features that capture the unique properties of each client's data distribution, such as attack type diversity and the balance between normal and malicious data instances, $P_k$ : performance metrics reflecting the effectiveness of each client's model, including indicators like validation accuracy and false positive rate, and $\Phi$: the meta-objective function that adjusts each $w_k$ to ensure a balanced representation of diverse data distributions, thereby minimizing the overall error and enhancing robustness. This optimization strategy enables a targeted and data-informed aggregation, allowing the model to handle IoT data heterogeneity more effectively while minimizing issues such as model bias and error due to imbalanced or limited datasets. To refine the adaptive weights $w_k$ for each client, a gradient-based optimization process is applied to iteratively minimize the meta-objective $\Phi$. This process recalculates $w_k$ by adjusting the weights at each iteration to maximize the overall model's generalization across diverse client datasets (Eq. 7) [41]:

$$w_k^{(t+1)} = w_k^{(t)} - \eta \cdot \frac{\partial \Phi}{\partial w_k} \qquad (7)$$

Where, $\eta$ represents the learning rate, which controls the step size for each weight adjustment, and $\frac{\partial \Phi}{\partial w_k}$ is the gradient of the meta-objective function $\Phi$ with respect to $w_k$, which guides the direction and magnitude of the update for optimal weight allocation. Through this gradient-based weight update, the model dynamically adjusts $w_k$ based on real-time changes in data characteristics and client performance metrics, ensuring the weights evolve adaptively as new data distributions emerge. This process not only strengthens the model's resilience to heterogeneity but also reduces the bias and error that typically arise from imbalanced datasets. As a result, the model maintains robust performance even in highly varied IoT environments, effectively capturing the nuances of diverse client data and reducing the likelihood of o
verfitting to any particular distribution. This gradient-based update enables the system to adaptively adjust each client's weight in response to evolving data distributions, improving the model's resilience to data heterogeneity.

*3.2.4. Enhanced Global Loss Function with Anomaly-Aware Regularization*

To effectively address data heterogeneity in IoT-based intrusion detection, an anomaly-aware regularization term is introduced within the global loss function. This additional term enables the global model to prioritize information from clients containing rare or underrepresented attack types, preventing the model from overfitting to majority-class data patterns and enhancing its sensitivity to diverse attack types commonly found in heterogeneous IoT networks. By focusing on both common and unique intrusion patterns across clients, the model improves its detection accuracy and resilience to varied data distributions (Eq. 8)

$$L_{global}(\theta_g) = \sum_{k=1}^{K} w_k \cdot L_k(\theta_k) + \alpha \cdot R(\theta_g) + \beta \cdot A(\theta_g) \qquad (8)$$

where:

- $L_k(\theta_k)$: Represents the local loss for each client k, which assesses the model's performance on the client's individual dataset and directly reflects the accuracy of intrusion detection at a local level.

- $R(\theta_g)$: A regularization term, commonly implemented as an L2-norm, which helps in constraining model parameters to prevent overfitting and promotes generalization, and it is defined as follows: (where $\|\theta_g\|_2^p$ L2 norm for model parameter). The regularization term aims to prevent overfitting by constraining the model parameters, ensuring that the model generalizes well to unseen data. By penalizing large parameter values, it promotes simplicity and robustness, which are critical for handling the diverse and heterogeneous data distributions in IoT environments.

$$R(\theta_g) = \frac{1}{2} \|\theta_g\|_2^p \tag{9}$$

- $A(\theta_g)$: An anomaly-based regularization term, which emphasizes contributions from clients with unique or rare attack types, ensuring these patterns are adequately represented in the global model. This term plays a crucial role in adapting the model to diverse and potentially rare intrusion scenarios across IoT devices.

$$A(\theta_g) = \sum_{i=1}^{n_k} w_i \cdot L_k(\theta_k) \tag{10}$$

- α and β: Hyperparameters that balance the contributions of $R(\theta_g)$ and $A(\theta_g)$, allowing for controlled influence of regularization and anomaly prioritization on the global model.

The hyperparameters α and β are carefully selected through grid search and cross-validation to optimize the trade-off between model generalization and sensitivity to rare attacks. Specifically, α regulates the regularization strength to prevent overfitting and ensure stable performance across diverse data, while β adjusts the emphasis on anomaly-aware regularization, enhancing detection of rare and underrepresented attacks. Experimental tuning revealed that increasing β improves the detection rate of rare attacks without significantly compromising accuracy on common attacks, whereas an appropriate α maintains overall robustness and reduces false positives. Thus, balancing these parameters is crucial to achieving strong performance in heterogeneous IoT intrusion detection scenarios. Incorporating the anomaly-aware regularization term $A(\theta_g)$ not only improves the model's adaptability to heterogeneous data but also reduces the likelihood of false positives and false negatives, which are critical metrics for intrusion detection accuracy. This targeted approach ensures that the global model is both robust and comprehensive, enabling it to respond effectively to the complexities and varied attack patterns inherent in IoT networks.

Finally, the provided pseudocode describes a Federated Learning (FL) system for intrusion detection in heterogeneous IoT environments, employing an Adaptive Meta-Aggregation mechanism. In this approach, an initial model with randomly initialized parameters $\theta_g$ is distributed by the central server to k IoT devices (clients), each possessing unique data characteristics. Upon receiving the global model, each client k trains its local model $M_k$ on its dataset $D_k$, resulting in updated parameters $\theta_k$ and calculating an adaptive weight $w_k$. This weight $w_k$ is derived from meta-features which, ensuring that clients with more diverse or high-quality data make a stronger impact on the aggregated model. The clients then send $\theta_k$ and $w_k$ to the central server, where a weighted aggregation of all $\theta_k$ values forms the updated global model $M_g$. This iterative process—model distribution, local training, and weighted aggregation continues across FL rounds, allowing the global model to adapt effectively to the data heterogeneity (Algorithm 1).

The computational complexity of the algorithm can be analyzed based on both the local model training and the global model aggregation. For each client k, the local model training involves iterating over the client's dataset $D_k$, and the complexity of training a local model $M_k$ can be expressed as O ($N_k \cdot D_k$), where $N_k$ is the number of iterations required for training the model on client k's data. Since the algorithm runs for T rounds and involves K clients, the total complexity of local model training for all clients is O (T· K· $N_k$· $D_k$). After training the local models, the global model is updated by aggregating the weighted parameters from each client. The aggregation step, which involves combining the local model parameters $θ_k$ from each client weighted by $w_k$ has a complexity of O (K ·θ), where θ is the number of parameters in the model. Therefore, the total complexity of the algorithm, considering both the local computations and global aggregation, is O (T· (K· $N_k$ · $D_k$+ K· θ)). This reflects the computational load for all clients across multiple rounds of model updates.

**Algorithm 1:** Training mechanism of Adaptive Meta-Aggregation Federated Learning for Intrusion Detection

1. Initialize global model parameters $θ$
2. for $t = 1$ to $T$ do
3.     for each client $k = 1$ to $K$ in parallel do
4.         Train local model $M_k$ on $D_k^{train}$ to obtain updated parameters $θ_k$
5.         Evaluate $M_k$ on $D_k^{val}$ to derive performance metric $P_k$
6.         Extract meta-features $X_k$ from $D_k^{train}$ (e.g., distribution metrics and class balance)
7.         Compute client weight $w_k$ based on $X_k$ and $P_k$
8.         Send $θ_k$, $P_k$, $X_k$, and $w_k$ to the Meta-Aggregator
9.     end for
10.     Aggregate $θ = \sum_{k=1}^{K} w_k \cdot θ_k$ to form the global model $M_g$
11.     Update $θ$ by optimizing the global loss function, incorporating anomaly-aware regularization
12.     Distribute the updated global model $θ$ to all clients for the next round
13. end for
14. Return optimized global model $θ^*$

4. **Evaluation Result**

In this section, the performance of the proposed method will be evaluated. All implementations and evaluations were performed on a machine equipped by an NVIDIA RTX 3080Ti GPU with 8GB of RAM. This setup was powered by an Intel Core i9-7700X processor with a base frequency of 3.50 GHz and a maximum turbo frequency of 4.00 GHz, supported by 128 GB of DDR4 RAM. The operating system used was Ubuntu 24.04, and the development environment included JupyterLab 3.6.7 IDE alongside Python. For the implementation of the proposed approach based on the PyTorch library was used, which provides a simple and intuitive API for building FL models [42].

Using five metrics Accuracy, Precision, F1-score, Recall, False Positive Rate (FPR), and True Positive Rate (TPR) the suggested method has been assessed. True Positives (TP), True Negatives (TN), False Positives (FP), and False Negatives (FN) parameters are taken into account when calculating these measures. A detailed description of the evaluation's outcomes utilizing these measures will be provided in the sections that follow. The next part will cover the parameters of the suggested methodology, the dataset that was used, the findings evaluation, and a comparison with previous research.

$$\text{Accuracy} = \frac{TP + TN}{TP + TN + FP + FN} \quad (11)$$

$$\text{Precision} = \frac{TP}{TP + FP} \quad (12)$$

$$\text{Recall} = \frac{TP}{TP + FN} \quad (13)$$

$$F1 - \text{score} = 2 * \frac{\text{Recall} * \text{precision}}{\text{Recall} + \text{precision}} \quad (14)$$

$$\text{False Positive Rate (FPR)} = \frac{FP}{FP+TN} \quad (15)$$

$$\text{True positive Rate (TPR)} = \frac{TP}{FN+TP} \quad (16)$$

*3.2. Experimental setting and parameter analysis*

In the proposed method, a CNN model has been used to evaluate the model. In this paper, the CNN architecture consists of 6 convolutional layers with 3 × 3 kernels, plus stride=1, padding=1, 6 pooling layers with 2 × 2 kernels, plus stride=1, and padding=1, as well as 2 fully connected layers. The model weights have been trained using the Adam optimizer. Batch normalization with a batch size of 128 is applied in this study. The learning rate is set to a constant value of 0.01 for the server, and 0.01 for the clients in the first round, with the addition of an epsilon value in each training round. Considering a total of 10 rounds for training (Rounds=10), the model is trained for 50 epochs in each training scenario. The reason for choosing 50 epochs is that in this scenario, the accuracy starts to decrease around epoch 20, so a total of 50 epochs is set for each case, despite the convergence of the other cases. The number of clients is set to 10. Additionally, an 80-20 ratio is defined between the training and the test sets (Table 3).

Table 3: Simulation parameters.

| Parameter | Value/Description |
|---|---|
| Model Architecture | CNN (6 convolutional layers, 3×3 kernels) |
| Pooling Layers | 6 layers, 2×2 kernels |
| Fully Connected Layers | 2 layers |
| Optimizer | Adam Optimizer |
| Batch Size | 128 |
| Learning Rate (Server) | 0.01 (constant) |
| Learning Rate (Clients) | 0.01 (constant in round 1, with epsilon) |
| Number of Training Rounds | 10 |
| Epochs per Round | 50 |
| Clients | 10 |
| Train-Test Split | 80-20 Ratio |
| A | 0.5 |
| B | 0.3 |
| $\gamma$ | 0.7 |
| $\delta$ | 0.3 |
| H | 0.01 |
| $\Lambda$ | 0.6 |

*3.3. Dataset*

To comprehensively evaluate the proposed AMAFed framework, three benchmark IoT intrusion detection datasets were employed: ToN-IoT, N-BaIoT, and BoT-IoT. These datasets represent diverse real-world IoT network scenarios with varying traffic patterns, device types, and attack distributions, enabling a thorough assessment of model generalization in heterogeneous environments.
- **ToN-IoT Dataset:** the ToN-IoT dataset [27] is a next-generation IoT and Industrial IoT (IIoT) dataset designed to capture and analyze heterogeneous data sources, including telemetry, network traffic, and system logs. It was developed to evaluate the performance of advanced cybersecurity applications leveraging AI and ML. The dataset is provided in CSV format and includes labeled entries categorized as either normal behavior or specific attack types. Table 4 summarizes the number of records per attack type and normal class. Data sources encompass diverse IoT devices and network configurations, ensuring a high degree of heterogeneity in both benign and malicious traffic patterns.
- **N-BaIoT Dataset:** the N-BaIoT dataset [28] consists of real network traffic captured from nine commercial IoT devices compromised by Mirai and Bashlite botnets. The traffic includes both benign activity and malicious patterns such as DDoS, UDP flooding, and TCP flooding. Features are extracted from raw packet captures, producing numerical representations suitable for ML-based intrusion detection. The dataset's heterogeneity arises from differences in device types, network behavior, and attack intensities, making it a valuable benchmark for evaluating IDS performance in distributed, device-diverse IoT environments.
- **BoT-IoT Dataset:** the BoT-IoT dataset [29] was generated in a controlled Cyber Range Laboratory environment at UNSW Canberra. It contains a mix of normal traffic and simulated botnet activities, including DDoS, DoS, OS and Service Scan, Keylogging, and Data Exfiltration attacks. Attacks are further organized by protocol type, providing detailed granularity for training and testing intrusion detection systems. The dataset is particularly challenging due to its high imbalance between benign and malicious samples and its coverage of a wide range of attack vectors.

By incorporating these three datasets, the evaluation covers a broad spectrum of IoT network heterogeneity, ensuring that the proposed method is tested against variations in traffic volume, device diversity, and attack complexity. This multi-dataset evaluation provides stronger evidence of AMAFed's generalization capabilities in real-world IoT scenarios.

*3.4. Result*

This section presents a thorough evaluation of the proposed method. The confusion matrix shown in Figure 3 clearly illustrates the model's exceptional ability to accurately classify a wide range of attack types within the heterogeneous ToN-IoT dataset. Precision, recall, and F1-scores consistently exceed 99% across most classes, demonstrating robustness despite challenges arising from the dataset's heterogeneity, such as imbalanced attack distributions and varying complexities. Particularly low false positive and false negative rates ensure reliable detection in critical categories including DoS, DDoS, and Scanning attacks, which constitute the majority of the dataset. Figure 4 demonstrates the performance of the proposed intrusion detection system in accurately identifying various attack types from the TON_IOT dataset. Each bar represents an attack category, with the corresponding accuracy percentage clearly indicated above. The results highlight the system's exceptional accuracy, exceeding **99%** for most attack types, including "DoS" (99.7%), "DDoS" (99.6%), and "Normal" traffic (99.6%). Notably, the system achieves the highest accuracy of **99.8%** for detecting "Scanning" attacks, showcasing its effectiveness in identifying network reconnaissance attempts. Despite the overall high performance, minor variations are observed for certain categories, such as "Ransomware" (98.5%) and "MITM" (98.7%), which slightly lag behind other attack types. These differences may be attributed to the inherent complexity or limited representation of these attacks in the dataset. Nevertheless, the proposed system maintains consistent accuracy across all attack categories, demonstrating its robustness and adaptability to diverse IoT environments. This strong performance underscores the effectiveness of the system's adaptive Meta-Aggregator and its ability to handle heterogeneous data. The results validate the capability of the intrusion detection system to generalize across varying attack patterns while maintaining minimal false positives and negatives, making it highly reliable for real-world IoT networks.

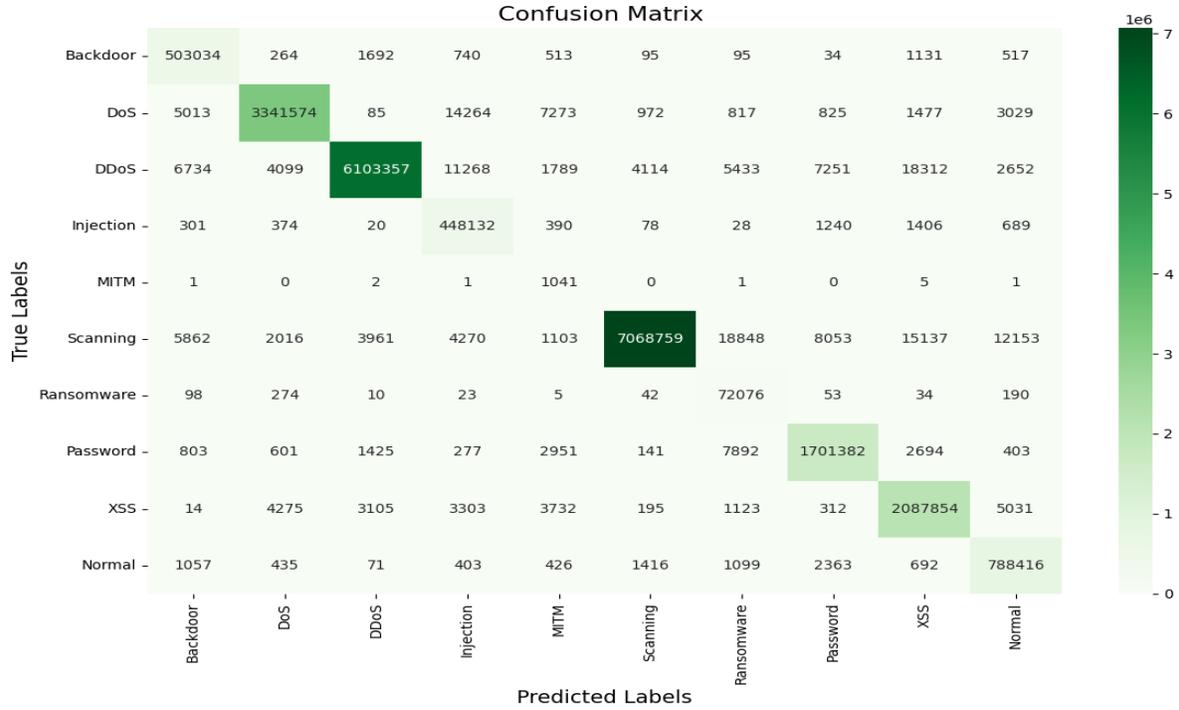

Figure 3: The confusion matrix of TON-IoT

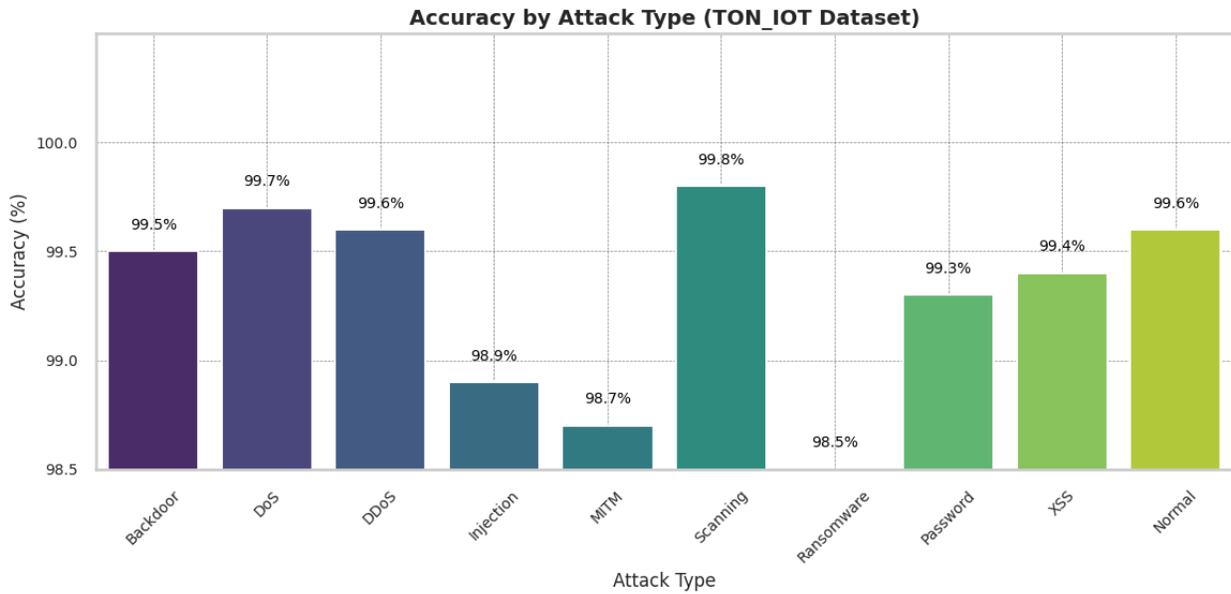

Figure 4: Evaluation of attack types

The ROC curve of the proposed model shows in figure 5, which its ability to effectively handle data heterogeneity in IoT environments. The steep initial rise indicates a high True Positive Rate (TPR) at low False Positive Rates (FPR), showcasing the model's strong detection capability. The curve maintains a consistent upward trajectory, reflecting its reliability across diverse client data distributions. The near-optimal TPR at higher FPR values highlights the effectiveness of the Meta-Aggregator in leveraging adaptive weights, ensuring accurate and generalized intrusion detection.

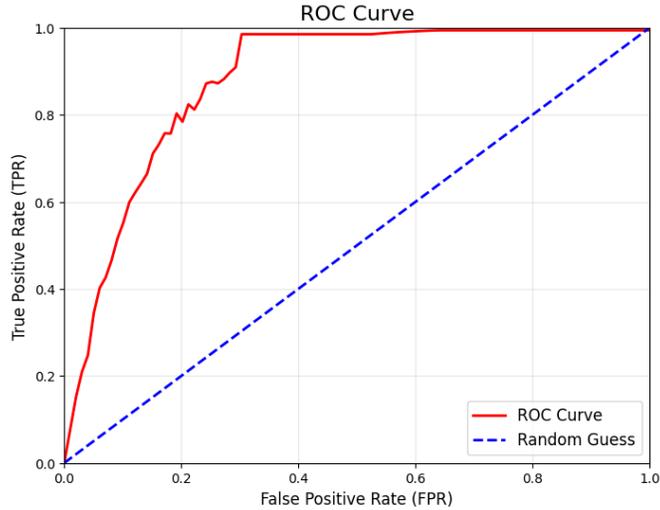

Figure 5: ROC curves of the proposed method for ToN-IoT

The evaluation results of the proposed approach for clients using the Accuracy, Precision, Recall, F1-score, and FPR metrics for the ToN-IoT dataset are presented in Figure 6, across 10 training rounds on the IoT devices in the ToN-IoT dataset. Each curve represents one of the 10 clients participating in the federated learning setup, illustrating the model's adaptability and consistent performance across heterogeneous IoT devices. The accuracy plot demonstrates a remarkable improvement for all clients, with scores quickly exceeding 99% after the third round and converging to nearly 99.8% by the final round. This highlights the exceptional learning capacity of the model in identifying both benign and malicious activities. Similarly, the precision graph reflects a high ability to reduce false positives, with clients achieving scores consistently above 99.5% from the fifth round onward. In terms of recall, the proposed method shows its strength in detecting true positives effectively. All clients surpass 99.2% recall by the sixth round, indicating the robustness of the system in identifying potential threats with minimal false negatives. The F1-score, which balances precision and recall, follows a similar trend, reaching 99.7% for most clients by the eighth round. This high level of consistency across all metrics validates the reliability of the proposed model. The results clearly underscore the effectiveness of the federated learning and meta-learning enhancements used in the proposed method. The steady and synchronized improvements across all clients demonstrate the robustness of the clustering strategy in addressing the heterogeneity of IoT devices. These findings highlight the potential of the proposed system as a scalable, efficient, and highly accurate solution for intrusion detection.

To strengthen the evaluation and make heterogeneity explicit, this part conducts a focused robustness study of AMAFed on ToN-IoT under non-IID, label-skewed client data. In label-skew, each client observes a different class mix some predominantly normal traffic, others dominated by specific attacks. The severity of this skew is controlled by a Dirichlet concentration parameter α ∈ {0.1,0.3,1.0}, where smaller α indicates heavier imbalance and therefore a harder setting. All training details remain exactly as defined earlier so that any changes in behavior can be attributed to the aggregation strategy under skew rather than to extra capacity or tuning. Robustness is first examined through convergence efficiency. For each round, the mean F1-score across participating clients is tracked, and a Rounds-to-Target criterion is reported. Robustness is then assessed in terms of cross-client stability at the final round. Beyond the overall mean, the dispersion of per-client F1-score is summarized together with two anchors that matter in practice: the worst-client F1-score and the 5th-percentile F1-score. High values for these anchors, accompanied by a tight spread, indicate that performance is consistent across clients and that weaker clients are not being sacrificed to raise the global average an essential property for intrusion detection where every device must remain adequately protected .Finally, robustness is quantified on the weak tail of clients. Here, "tail" refers to the small group with the lowest F1-score. The empirical cumulative distribution function (ECDF) of final-round per-client F1-score is used to make this explicit for any threshold x, the ECDF gives the fraction of clients with F1 ≤x. Figures 1-3 report the robustness analysis of AMAFed under label-skew non-IID splits on ToN-IoT.

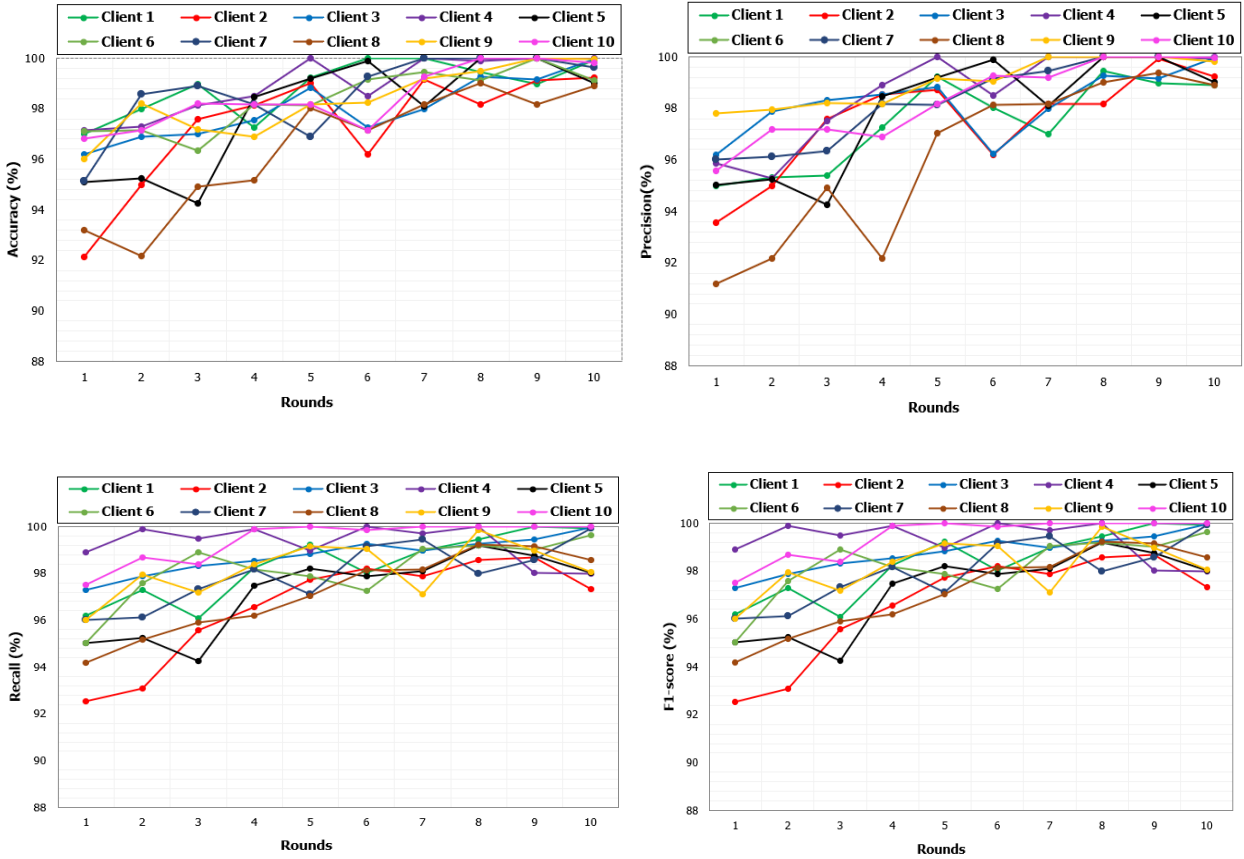

Figure 6: Evaluation Accuracy, Precision, Recall and F1-score metrics of training rounds in proposed method on ToN-IoT dataset.

Figure 7 shows the learning curves (F1 vs. rounds) for AMAFed and FedAvg across the three α levels. AMAFed climbs faster and saturates higher in every case. Under strong and moderate skew (α=0.1,0.3), AMAFed reaches the % 0.98 regime by rounds 7–8, while FedAvg remains below that level within the same budget. Under milder skew (α=1.0), AMAFed maintains a stable plateau around 0.99%, whereas FedAvg peaks lower. This pattern is consistent with AMAFed's adaptive meta-aggregation, which up-weights informative clients and dampens conflicting updates.

Figure 8 illustrates the final-round distribution of per-client F1-SCORE. AMAFed exhibits tighter IQRs and higher lower whiskers than FedAvg for all α's, indicating less spread across clients and stronger guarantees for weaker sites. Medians for AMAFed cluster near 0.98–0.99, while FedAvg centers around %0.93 (α=0.1), %0.94 (α=0.3), and % 0.96 (α=1.0). Higher worst-client and 5th-percentile levels confirm that AMAFed improves fairness rather than trading off weak clients to raise the mean. Figure 9 shows the ECDF of final-round per-client F1-score (example α=0.1) to make the weak-client tail explicit. For any threshold x, the curve reports the fraction of clients with F1 ≤x. AMAFed's ECDF is right-shifted, concentrating clients in the 0.98–0.99 band; FedAvg's curve lies around 0.92–0.96, implying a much larger share below operational cutoff (F1-SCORE<0.95). This demonstrates substantially lower tail risk for AMAFed. Across all heterogeneity levels, AMAFed demonstrates faster and higher convergence, tighter cross-client dispersion with stronger worst-client guarantees, and a substantially smaller weak-client tail than FedAvg under the same budget evidence of robust behavior under realistic non-IID conditions.

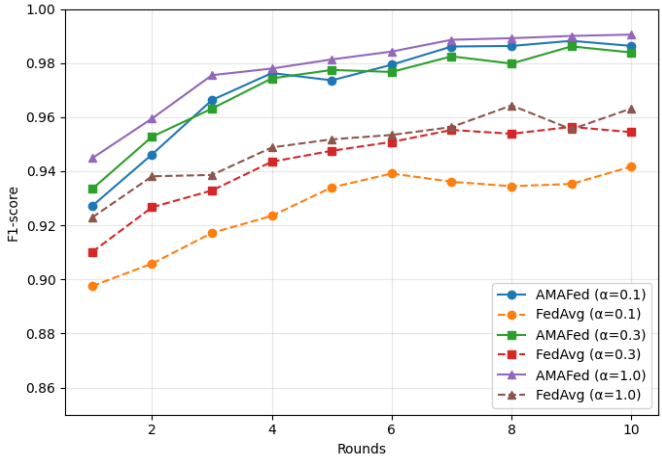
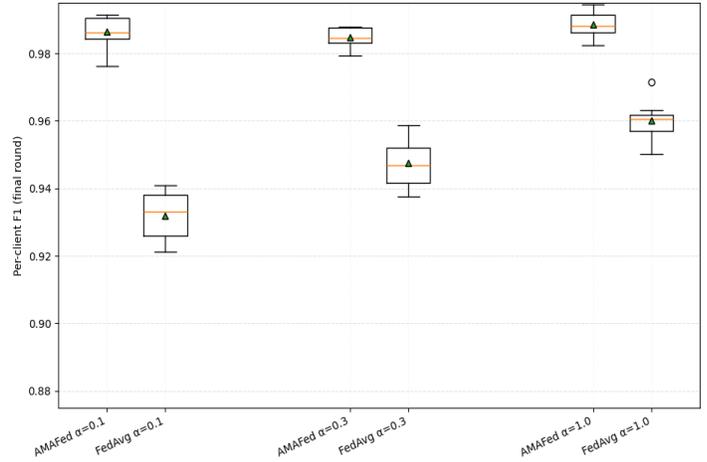

Figure 7. F1-score vs. rounds under label-skew heterogeneity.   Figure 8. Final-round per-client F1-score across α.

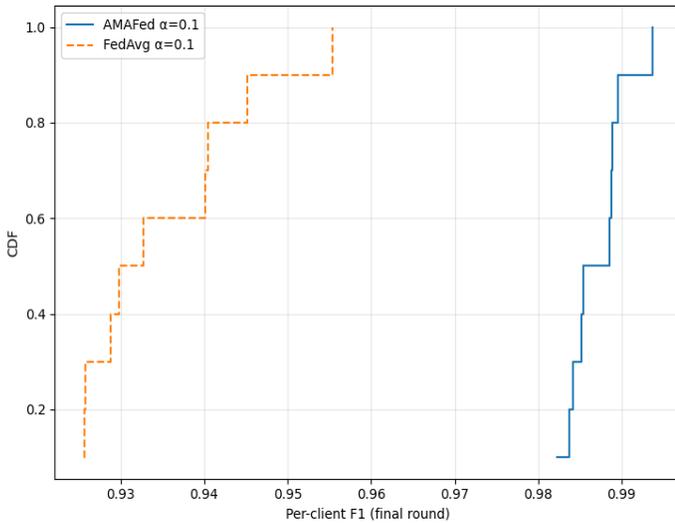
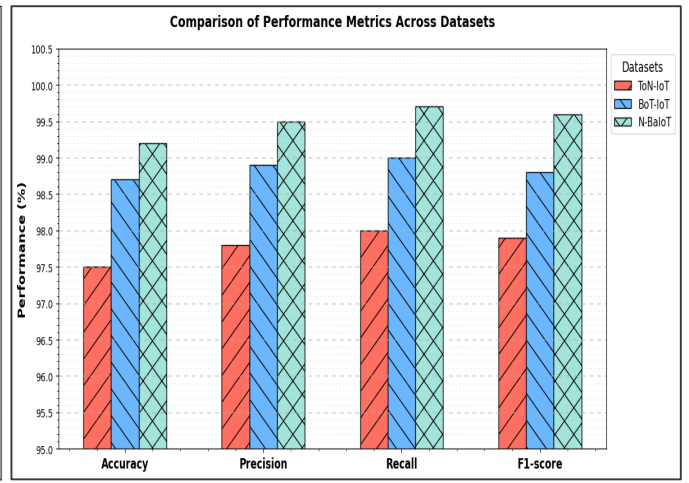

Figure 9. ECDF of per-client F1 (tail risk; α=0.1).   Figure 10: Evaluation of the proposed method on different datasets.

To evaluate the performance of the proposed approach, two additional datasets, namely N-BaIoT and BoT-IoT, were utilized [28] [29]. The results of experiments on these two datasets are presented in Table 7. The evaluation results indicate a good performance of the proposed approach on these different datasets. As shown in the table, the proposed approach achieved an accuracy of 98.12% on the BoT-IoT dataset and 98.88% on the N-BaIoT dataset. The F1-score value for the BoT-IoT dataset is 97.95%, indicating a good performance of the proposed approach in detecting attacks. Furthermore, for the N-BaIoT dataset, the obtained F1-score value is 98.01, which is higher than the F1-score value for the first dataset. Other evaluation metrics are also provided in the table 4.

Table 4: Performance evaluation of the proposed approach on other datasets (%).

| Dataset | Accuracy | Precision | Recall | F1-score |
|---|---|---|---|---|
| BoT-IoT[29] | 98.12 | 98.92 | 99.25 | 97.95 |
| N-BaIoT[28] | 99.88 | 99.11 | 98.91 | 98.01 |

In Figure 10, the evaluation charts for the metrics Accuracy, Precision, Recall, and F1-score are depicted for three datasets: ToN-IoT, BoT-IoT, and N-BaIoT. As observed in the figure, the proposed approach has achieved good performance for all three datasets. The best performance of the proposed approach is seen in terms of the Accuracy and Precision metrics for the N-BaIoT dataset. The BoT-IoT dataset has exhibited higher performance in terms of the Recall metric compared to the other two datasets. In intrusion detection for IoT attacks, achieving a high F1-score is crucial as it ensures that the model can effectively identify attacks while minimizing false alarms. A higher value has been obtained for the heterogeneous dataset ToN-IoT.

Table 5. Performance evaluation of the proposed approach on other datasets (%).

| Research | Accuracy (%) | F1-score (%) | Recall (%) | Precision (%) |
|---|---|---|---|---|
| Abdel-Basset et al. [43] | 94.85 | 93.13 | 93.09 | 93.17 |
| Booij et al. [20] | 97.82 | 92.12 | - | - |
| Wang et al. [44] | 97.06 | 96.94 | 96.67 | 97.23 |
| Proposed method | **99.12** | **98.50** | **98.52** | **98.33** |

### 4.4. Comparing with other methods

Table 5 compares the AMAFed approach with other methods in terms of accuracy, recall, precision, and F1-score metrics. The proposed method achieves superior results compared to previous works, particularly when considering the challenges posed by the heterogeneous nature of the dataset. In contrast to earlier studies, which primarily evaluated performance on the ToN-IoT dataset, the proposed approach effectively addresses the data heterogeneity issue by leveraging advanced techniques in federated learning. This allows for a more accurate evaluation and better generalization across varied data distributions, which is a crucial aspect of real-world IoT environments. While prior methods, such as those by Abdel-Basset et al. [43], Booij et al. [20], and Wang et al. [44], demonstrated promising results, the approach presented in this study offers a more robust and consistent performance across multiple evaluation metrics. By overcoming the limitations of data heterogeneity, the proposed method not only achieves higher accuracy but also ensures enhanced precision, recall, and F1-score, further solidifying its effectiveness in IoT-based intrusion detection systems. The improvements in these metrics highlight the method's ability to perform more reliably in complex and diverse IoT scenarios, positioning it as a more advanced and practical solution for intrusion detection.

## 5. Conclusion & Future Work

This study addresses the critical challenge of data heterogeneity in IoT environments by proposing the Adaptive Meta-Aggregation Federated Learning (AMAFed) framework. The method leverages a dynamic weighting mechanism to prioritize contributions from IoT devices with diverse and high-quality datasets, effectively mitigating the limitations of traditional intrusion detection systems. By employing an adaptive meta-aggregator and a hybrid loss function, the proposed approach enhances sensitivity to rare attack patterns while maintaining robust overall performance. Extensive eᵛvaluations on benchmark datasets, including ToN-IoT, N-BaIoT, and BoT-IoT, demonstrate the efficacy of the method, achieving detection accuracy exceeding 98% and significantly outperforming existing approaches. These results validate the potential of AMAFed as a scalable and adaptable solution for securing heterogeneous IoT networks against diverse intrusion threats. For future work, the focus will be on further enhancing the scalability of the framework by integrating energy-efficient aggregation mechanisms suitable for resource-constrained IoT devices. Additionally, the exploration of transfer learning techniques can improve model generalization across unseen attack patterns and new IoT devices. Another promising direction is the incorporation of reinforcement learning to enable real-time adaptation to rapidly evolving data distributions and attack strategies. By addressing these aspects, the proposed method can be refined to achieve greater robustness and reliability, ensuring comprehensive security for next-generation IoT ecosystems.

**Funding**: No funding
**Ethic declaration**: Not applicable
**Author contributions**: Saadat Izadi: Implementing, simulating and writing the draft, and Mahmood Ahmadi: Reviewing and checking the final draft.
**Data availability:** Data will be available by request.
**Competing Interest declaration**: The authors declare no competing interests.